\RequirePackage{fix-cm}
\documentclass[smallextended]{svjour3}       
\smartqed  
\usepackage{graphicx}
\usepackage{color}
\usepackage{amsmath}
\usepackage{amsfonts}
\usepackage{amssymb}
\newcommand{\ket}[1]{|#1\rangle}
\newcommand{\bra}[1]{\langle #1|}
\newcommand{\Tr}{\textrm{Tr}}

\journalname{Quantum Information Processing}
\begin{document}

\title{Conceptual aspects of geometric quantum computation}


\author{Erik Sj\"oqvist \and Vahid Azimi Mousolou \and Carlo M. Canali}

\institute{Erik Sj\"oqvist \at 
Department of Physics and Astronomy, Uppsala University, Box 516, Se-751 20 Uppsala, Sweden \\
Tel.: +46-73-0245822\\
\email{erik.sjoqvist@physics.uu.se} 
\and
Vahid Azimi Mousolou \at
Department of Mathematics, Faculty of Science, University of Isfahan, 
Box 81745-163 Isfahan, Iran
\and 
Carlo M. Canali \at 
Department of Physics and Electrical Engineering, Linnaeus University,
391 82 Kalmar, Sweden}

\date{Received: date / Accepted: date}

\maketitle

\begin{abstract}
Geometric quantum computation is the idea that geometric phases can be used to 
implement quantum gates, i.e., the basic elements of the Boolean network that forms 
a quantum computer. Although originally thought to be limited to adiabatic evolution, 
controlled by slowly changing parameters, this form of quantum computation can as well 
be realized at high speed by using nonadiabatic schemes. Recent advances in quantum 
gate technology have allowed for experimental demonstrations of different types of 
geometric gates in adiabatic and nonadiabatic evolution. Here, we address some 
conceptual issues that arise in the realizations of geometric gates. We examine the 
appearance of dynamical phases in quantum evolution and point out that not all 
dynamical phases need to be compensated for in geometric quantum computation. 
We delineate the relation between Abelian and non-Abelian geometric gates and 
find an explicit physical example where the two types of gates coincide. We identify 
differences and similarities between adiabatic and nonadiabatic realizations of 
quantum computation based on non-Abelian geometric phases. 
\end{abstract}

\section{Introduction}
More than 15 years ago, Zanardi and Rasetti \cite{zanardi99} demonstrated that a 
non-Abelian (matrix-valued)  geometric phase of a generic  pair of adiabatic loops in 
parameter space is sufficient to execute any information processing on a quantum 
computer. This all-geometric form of quantum computation has since then attracted 
considerable interest because of its potential robustness to parameter noise 
\cite{pachos01} and its conceptually appealing relation to the geometric 
description of quantum systems \cite{lloyd01}. 

The realization of a quantum computer requires a certain sequence of quantum gate 
operations acting on a set of two-level systems (qubits). The goal of geometric quantum 
computation is to implement each of these gates by using geometric phases only. Such 
gates may be realized by using geometric phases arising in adiabatic 
\cite{zanardi99,ekert00,duan01,faoro03,solinas03a} or nonadiabatic 
\cite{xiang-bin01,zhu02,zhu03a,zhu03b,sjoqvist12} evolution. Experimentally, geometric 
gates have been performed in nuclear magnetic resonance \cite{jones00,du06,feng13}, 
ion traps \cite{leibfried03,toyoda13}, superconducting qubits \cite{abdumalikov13}, and 
solid-state systems \cite{tian09,arroyo-camejo14,zu14}. Thus, geometric quantum computation 
is a well-established approach to quantum gate architecture. 

Geometric quantum computation involves adiabatic or nonadiabatic, as well as Abelian 
or non-Abelian characteristics of the underlying quantum evolution. In each of the proposed 
schemes cited above, a particular combination of these characteristics has been considered. 
The different combinations are associated with certain conceptual issues, related to the physical 
nature of the time evolution as well as to the underlying geometric structure of the state space 
of the qubits. The aim of the present work is to shed light on some of 
these conceptual issues that arise in the realizations of geometric gates. 

A central element in all schemes for geometric quantum computation is to develop methods 
to make dynamical phases irrelevant in order to achieve purely geometric transformation 
effects. However, in a given physical realization of a quantum gate, there might be different 
forms of dynamical phases involved, of which not all are relevant to the gate operation. 
Thus, such dynamical phases can be allowed for without affecting the geometric nature 
of the gates. The aim of the analysis in Sec.~\ref{sec:dp} is to identify these dynamical 
phases and to demonstrate why they are harmless in geometric quantum computation. 

Zhu and Wang (ZW) \cite{zhu02,zhu03a} have pointed out that nonadiabatic Abelian 
geometric phases \cite{aharonov87} are sufficient for universal all-geometric quantum 
computation, despite the fact that such phases are U(1) and therefore commuting. The trick 
is to consider geometric phase shift gates in different bases in order to make the gates 
noncommuting. The alternative route proposed in Ref. \cite{sjoqvist12} to achieve universality  
is based on using nonadiabatic non-Abelian geometric phases \cite{anandan88}. Here, the gates 
are obtained by moving the computational system, which spans a subspace of a larger 
Hilbert space, around a loop, resulting in non-Abelian geometric phases. Although being 
conceptually very different, the two approaches achieve exactly the same: an all-geometric 
set of gates based on nonadiabatic evolution. This curious fact raises the question of whether 
there is any relation between the two approaches. In Sec.~\ref{sec:zwhqc}, we address this 
issue by demonstrating that the gates proposed in Ref.~\cite{sjoqvist12} can in fact be 
interpreted as ZW gates. 

Geometric gates can be generated either by adiabatic or by nonadiabatic evolution. These 
two types of gates have both differences as well as similarities. The purpose of 
Sec.~\ref{sec:a_vs_na} is to examine these differences and similarities in the case 
of gates based on non-Abelian geometric phases. Specifically, we examine the role 
of the run-time, the exactness of the gates, the role of the control parameters, 
and the interpretation of the loops that generate the geometric phases. As an example 
of adiabatic versus nonadiabatic evolution, we apply the general findings to the cases 
of the tripod and the $\Lambda$ schemes, which are realizations of adiabatic 
\cite{duan01} and nonadiabatic \cite{sjoqvist12} non-Abelian geometric gates, 
respectively.  

\section{Dynamical phases accompanying geometric phase shift gates}
\label{sec:dp}
A central element in all schemes for geometric quantum computation is the elimination 
of dynamical phase effects on the gate operation. This can be achieved in different ways 
depending on the main characteristics involved in the realization of a given geometric 
gate. It may involve specific paths in state space \cite{solinas03b,tian04,ota09} or 
tuning applied fields in certain ways \cite{zhu02,zhu03a,zhu05}. 

In some of these schemes, there may appear phases that do not affect the gate operation 
but nevertheless have a purely dynamical character. The purpose of this section is to clarify 
the distinction between these `harmless' dynamical phases from those that are necessary 
to compensate for in order to make the gates geometric. We limit the discussion 
to the case of Abelian geometric phases, while keeping in mind that similar arguments 
apply also to certain non-Abelian settings, such as that of Ref. \cite{mousolou14}. 

To delineate the basic idea of an Abelian geometric quantum gate, consider the simplest 
nontrivial case where a qubit evolves under the unitary evolution $U(t,0)$, $t\in [0,\tau]$. 
Let $\ket{0}$ and $\ket{1}$ be the eigenstates of $U(\tau,0)$ with corresponding eigenvalues 
$e^{i\varphi_0}$ and $e^{i\varphi_1}$. This defines the phase shift gate 
\begin{eqnarray} 
U(\tau,0):\ket{x} \mapsto e^{i\varphi_x} \ket{x}, \ x=0,1, 
\end{eqnarray}
which is nontrivial provided the relative phase $\varphi_0 - \varphi_1$ is not an integer 
multiple of $2\pi$. 

In general, $\varphi_x$ can be decomposed into a sum of a dynamical phase $\delta_x$ 
and a geometric phase $\gamma_x$, i.e., $\varphi_x = \delta_x + \gamma_x$. 
Explicitly, these read ($\hbar = 1$ from now on) 
\begin{eqnarray}
\delta_x & = &  - i \int_0^{\tau} 
\bra{x} U^{\dagger}(t,0) \dot{U} (t,0) \ket{x} dt 
\nonumber \\ 
 & = & - \int_0^{\tau} 
\bra{x} U^{\dagger}(t,0) H(t) U(t,0) \ket{x} dt , 
\nonumber \\ 
\gamma_x & = & \arg \bra{x} U(\tau,0) \ket{x} + i \int_0^{\tau} 
\bra{x} U^{\dagger}(t,0) \dot{U} (t,0) \ket{x} dt 
\nonumber \\ 
 & = &
\left( x - \frac{1}{2} \right) \Omega , 
\label{eq:relativephases}
\end{eqnarray}
where $H(t)$ is the Hamiltonian and $\Omega$ is the solid angle enclosed on the Bloch 
sphere. If $\delta_0 - \delta_1 = 0\  (\textrm{mod}\ 2\pi)$, then $U(\tau,0)$ defines a 
geometric phase shift gate, 
\begin{eqnarray}
U(\tau,0) \equiv {\textrm{U}}_g : \ket{x} \mapsto e^{i(x-\frac{1}{2}) \Omega} \ket{x} . 
\end{eqnarray}
${\textrm{U}}_g$ can be implemented either by using parallel transport \cite{solinas03b,tian04,ota09}, 
i.e., by imposing the condition $\bra{x} U^{\dagger}(t,0) \dot{U} (t,0) \ket{x} = 0$ 
throughout the evolution, or parameter tuning \cite{zhu02,zhu03a,zhu05} such that 
$\delta_x = {\textrm{integer}} \times 2\pi$, to remove the dynamical phases. 

\begin{figure}[t]
\centering
\includegraphics[width=85mm,height=64mm]{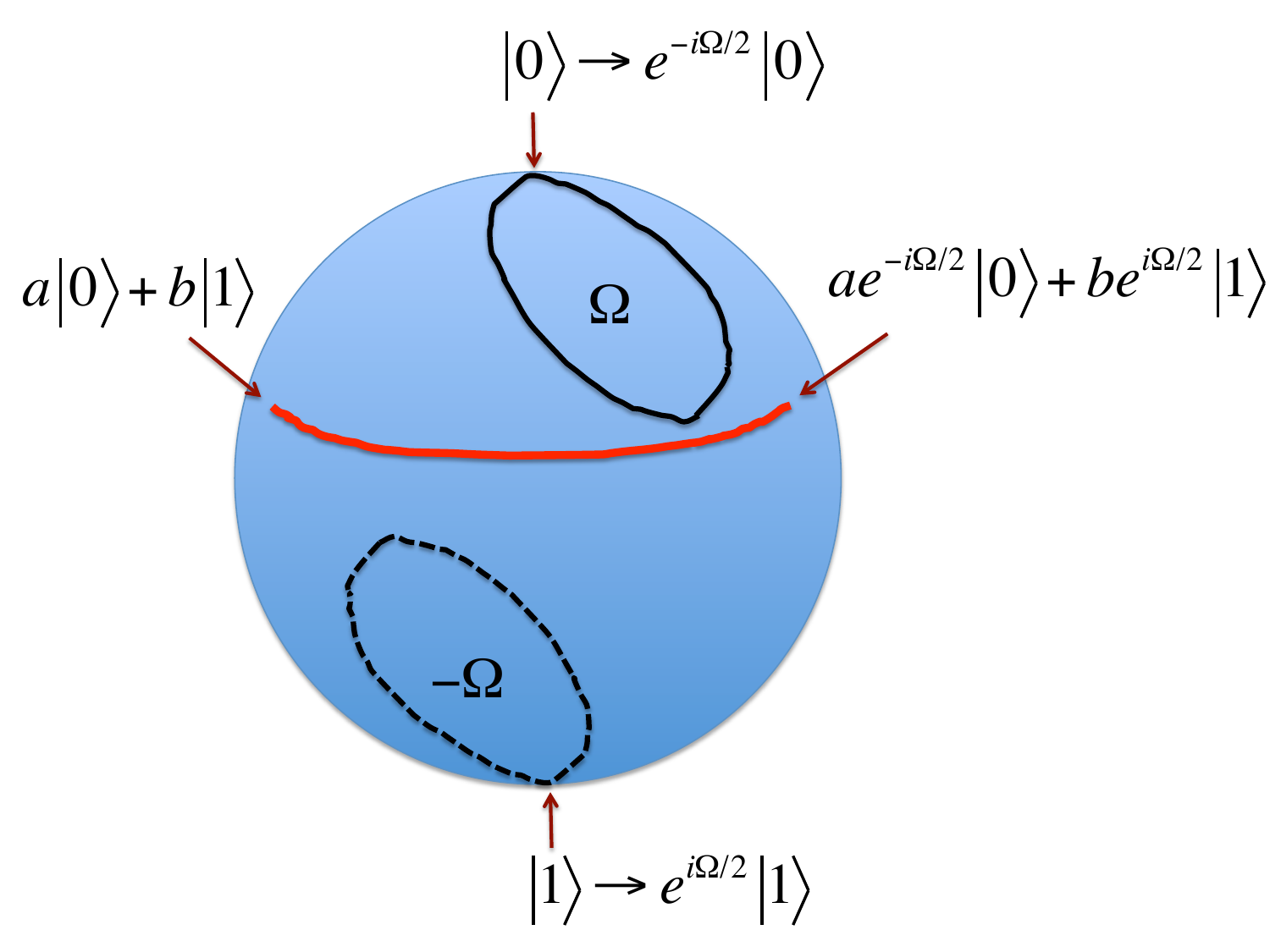}
\caption{Paths on the Bloch sphere, along which the states $\ket{0},\ket{1}$ and a general 
state $a\ket{0}+b\ket{1}$ evolve with $U(t,0)$, respectively. The states 
$\ket{0},\ket{1}$ traverse a closed path, while a general linear combination of these states 
does not need to go around a loop. The gate $\textrm{U}_g$ corresponds to the loops and global 
nonlinear dynamical and geometric phases $\Delta$ and $\Gamma$ correspond 
to the open path.} 
\label{fig:paths}
\end{figure}

Now, the action of ${\textrm{U}}_g$ on an arbitrary qubit state $\ket{\psi} = a\ket{0} + b\ket{1}$ 
reads 
\begin{eqnarray} 
a\ket{0} + b\ket{1} \mapsto 
a e^{-i\Omega /2}\ket{0} + be^{i\Omega /2}\ket{1} ,  
\end{eqnarray}
which implies that although the computational basis states $\ket{0}$ and $\ket{1}$ evolve in cyclic 
fashion on $t\in [0,\tau]$, a general linear combination of these states does not necessarily traverse 
a closed path on the Bloch sphere in this time interval, see Fig.~\ref{fig:paths}. Associated with this 
state change, there is a dynamical phase $\Delta$, which takes the form 
\begin{eqnarray}
\Delta  & = & - i \int_0^{\tau} \bra{\psi} U^{\dagger} (t,0) 
\dot{U} (t,0) \ket{\psi} dt 
\nonumber \\ 
 & = & - \left( |a|^2 - |b|^2 \right) \int_0^{\tau} \bra{0} U^{\dagger}(t,0) H(t) U(t,0) 
\ket{0} dt 
\nonumber \\ 
 & & - 2 {\textrm{Re}} \left( ab^{\ast} \int_0^{\tau} \bra{1} U^{\dagger}(t,0) H(t) U(t,0) 
\ket{0} dt \right) , 
\label{eq:dyn}
\end{eqnarray}
where we have chosen the zero-point energy such that $\int_0^{\tau} \Tr H(t) dt = 0$ for 
convenience. 

Clearly, $\Delta$ is not necessarily an integer multiple of $2\pi$; a fact that may cause some 
doubts concerning the geometric nature of ${\textrm{U}}_g$. These doubts can however be 
removed by noting that $\Delta$ is a  global phase, being part of the Pancharatnam phase 
$\arg \bra{\psi} {\textrm{U}}_g \ket{\psi}$ \cite{pancharatnam56}, and that it is a nonlinear 
functional of the input state $\ket{\psi}$ \cite{blais03}, while ${\textrm{U}}_g$ is a linear 
transformation of the relative phase between the two computational basis states $\ket{0}$ 
and $\ket{1}$. For these reasons, it follows that $\Delta$ is not part of the gate operation, 
which in turn ensures that ${\textrm{U}}_g$ is fully geometric.  

We further note that 
\begin{eqnarray} 
{\textrm{U}}_g \ket{\psi} = e^{i\arg \bra{\psi} {\textrm{U}}_g \ket{\psi}} \ket{\psi^{\parallel}} , 
\end{eqnarray}
where $\ket{\psi^{\parallel}}$ is parallel to $\ket{\psi}$ in the sense of the Pancharatnam 
connection, i.e., $\bra{\psi} \psi^{\parallel} \rangle >0$. The global phase 
$\arg \bra{\psi} {\textrm{U}}_g \ket{\psi}$ can be decomposed into the dynamical 
phase $\Delta$ and the remainder $\arg \bra{\psi} {\textrm{U}}_g \ket{\psi} - \Delta$. 
The remainder is invariant under the gauge transformation $U(t,0)\ket{\psi} 
\mapsto e^{if(t)} U(t,0)\ket{\psi},$ $t\in [0,\tau]$, i.e., it is the global geometric phase 
$\Gamma$ of the state. Geometrically, $\Gamma$ is minus half the solid angle enclosed by 
loop consisting of the open path shown in Fig.~\ref{fig:paths} and a geodesic connecting 
its end-points $a\ket{0}+b\ket{1}$ and $a e^{-i\Omega/2}\ket{0} + be^{i\Omega/2}\ket{1}$. 
Clearly, just as $\Delta$, $\Gamma$ is a global phase and a nonlinear functional of the input 
state $\ket{\psi}$, and therefore irrelevant to the gate operation. 

The fact that two vectors in Hilbert space associated with a quantum system represent 
the same quantum state if their overlapping probability amplitude is unity, or in other words 
if they correspond to the same point in the projective Hilbert space, justifies the irrelevance 
of global phase factors to quantum gate operations. Thus, $\Delta$ and $\Gamma$ are 
irrelevant to the gate since they are part of the unobservable global phase 
$\arg \bra{\psi} {\textrm{U}}_g \ket{\psi}$.  

The irrelevance of global phase factors to phase shift gates is utilized to eliminate dynamical 
phases in adiabatic geometric quantum gates by using spin echo techniques as follows. A spin 
(for instance of a nuclei in an NMR quantum computer) is taken around a loop $C$ by a slowly 
varying magnetic field. This results in a geometric phase factor $e^{\mp i \Omega /2}$, $\Omega$ 
being the solid angle enclosed by $C$ and the sign depending on whether the spin is aligned 
or anti-aligned with the magnetic field. Spin echo is based on the sequence $C \rightarrow \pi 
\rightarrow C^{-1} \rightarrow \pi$, where $\pi$ is a rapid spin flip operation. This scheme results 
in a geometric gate transformation \cite{ekert00}
\begin{eqnarray} 
\ket{x} \mapsto e^{i\left[ \delta_0 + \delta_1 + 2(x-\frac{1}{2}) \Omega \right]} \ket{x} , 
\end{eqnarray}
where $\delta_x$ are the dynamical phases picked up by the two spin eigenstates $\ket{x}$. 
The dynamical phases appear as a global phase and are therefore irrelevant to the gate 
operation, just as the dynamical phase $\Delta$ is irrelevant to the nonadiabatic geometric 
phase shift gate discussed above. 

To sum up, dynamical phases can occur as relative ($\delta_x$) and global ($\Delta$) 
phases, given in Eqs. (\ref{eq:relativephases}) and (\ref{eq:dyn}), respectively. While the 
relative dynamical phases are necessary to cancel or compensate 
for in order to implement a geometric phase shift gate, we can allow for a nontrivial global 
dynamical phase in such a gate due to the unobservability of the global phase in quantum 
mechanics. 

\section{Zhu-Wang versus non-Abelian nonadiabatic GQC}
\label{sec:zwhqc}

The Zhu-Wang (ZW) scheme \cite{zhu02,zhu03a} is a method to achieve universal geometric 
quantum computation based on Abelian nonadiabatic geometric phases only. In the 
one-qubit case, the idea is to consider the phase shift gate $\ket{\psi_\pm} \rightarrow 
e^{\pm i\gamma} \ket{\psi_{\pm}}$ with respect to the orthonormal  basis states 
$\ket{\psi_{+}} = \cos \frac{\chi}{2} \ket{0} + i\sin \frac{\chi}{2} \ket{1})$ and $\ket{\psi_{-}} = 
i\sin \frac{\chi}{2} \ket{0} + \cos \frac{\chi}{2} \ket{1})$. The cyclic phases $\pm \gamma$ 
coincide with the geometric phases $\mp \Omega/2$ picked up by $\psi_{\pm}$ during the 
evolution ($\Omega$ is the solid angle enclosed on the Bloch sphere) provided the dynamical 
phases are eliminated either by  employing rotating driving fields with fine-tuned parameters 
\cite{zhu02,zhu03a,zhu05} or by driving the qubit along geodesics on the Bloch sphere by 
using composite pulses \cite{solinas03b,tian04,ota09}. With respect to the computational 
standard basis $\ket{0},\ket{1}$, the resulting geometric gate takes the form 
\begin{eqnarray}
U_g^{\textrm{ZW}} & = & \left( e^{-i \Omega /2} \cos^2 \frac{\chi}{2} + 
e^{i\Omega /2} \sin^2 \frac{\chi}{2} \right) \ket{0} \bra{0} 
\nonumber \\ 
 & & +\sin \chi \sin \frac{\Omega}{2} (\ket{1} \bra{0} - \ket{0} \bra{1}) 
\nonumber \\ 
 & & + \left( e^{-i\Omega /2} \sin^2 \frac{\chi}{2} + 
e^{i\Omega /2} \cos^2 \frac{\chi}{2} \right) \ket{1} \bra{1} . 
\end{eqnarray}
To see that $U_g^{\textrm{ZW}}$ is sufficient for universality, we note that a geometric phase 
shift gate $\ket{x} \rightarrow e^{ix\Omega} \ket{x}$, can be implemented by 
choosing $\chi = 0$ and the Hadamard $\ket{x} \rightarrow \frac{1}{\sqrt{2}} \left(
\ket{x} + (-1)^x \ket{x \oplus 1} \right)$ can similarly be implemented by choosing 
$\chi = \Omega = \frac{\pi}{2}$. These gates are known to be universal for a single qubit. 

Another method to achieve fast universal geometric quantum computation is based on 
nonadiabatic non-Abelian geometric phases \cite{anandan88}. In its simplest form, this 
is achieved in a three-level $\Lambda$ configuration. By choosing common pulse envelope 
of the two drive fields, which couple a two-dimensional ground-state manifold (qubit state space in 
this configuration) to an auxiliary excited state, the dynamical phases can be shown to vanish 
at all times in nonadabatic evolution of the ground-state space. This results in purely geometric 
unitary transformation on the computational subspace spanned by the two ground-state levels. 
This is the basic idea behind the recently proposed \cite{sjoqvist12} and experimentally 
implemented \cite{abdumalikov13,feng13,arroyo-camejo14,zu14} nonadiabatic non-Abelian 
geometric gates for quantum information processing. 

Here, we show how ZW geometric quantum computation can be implemented in 
a three-level $\Lambda$ system. The purpose is to clarify the relation between the 
ZW idea \cite{zhu02,zhu03a} and geometric quantum computation based on nonadiabatic 
non-Abelian geometric phases \cite{sjoqvist12}. In fact, it turns out that the two schemes  
lead to identical gates in the $\Lambda$ system, which implies that the gates based on 
non-Abelian geometric phases proposed in Ref. \cite{sjoqvist12} can be interpreted as ZW 
gates. 

\begin{figure}[t]
\centering
\includegraphics[width=90mm,height=60mm]{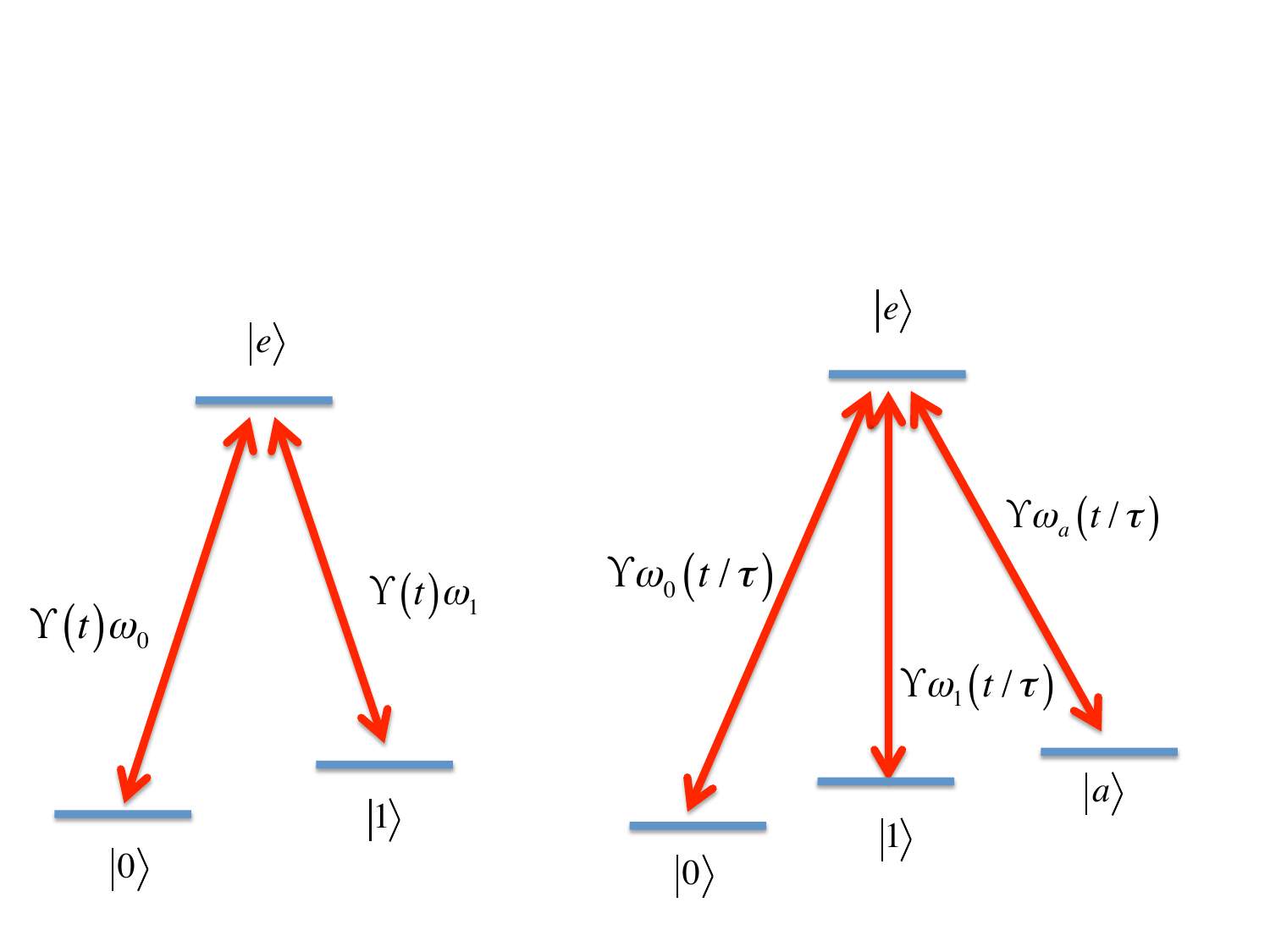}
\caption{$\Lambda$ (left panel) and tripod (right panel) system, in which an excited state 
$\ket{e}$ is coupled to two and three quasi-degenerate ground state levels, respectively. 
The $\omega_j$'s are complex-valued coupling parameters that can be controlled 
experimentally. The $\Lambda$ system is used to implement nonadiabatic geometric 
gates by using the same envelope function $\Upsilon (s)$ and time-independent 
$\omega_0,\omega_1$ for the pulses; the tripod system is used to implement adiabatic 
geometric gates by slowly varying the control parameters $\omega_0,\omega_1,\omega_a$ 
around a loop in parameter space.} 
\label{fig:tripodlambda}
\end{figure}

The basic zero-detuned Hamiltonian structure of the scheme in Ref. \cite{sjoqvist12} takes 
the form (see left panel of Fig.~\ref{fig:tripodlambda})
\begin{eqnarray}
 & & H (s) = \Upsilon (s) \left( \omega_0 \ket{e} \bra{0} +
\omega_1 \ket{e} \bra{1} + {\textrm{h.c.}} \right) 
\end{eqnarray}
with $s\in [0,1]$. The evolution of the qubit subspace spanned by $\ket{0}$ and $\ket{1}$ under 
this Hamiltonian results in the single-qubit gate 
\begin{eqnarray}
U (C_{{\bf n}}) & = & P_c e^{-i\tau \int_0^1 H(s) ds} P_c = {\bf n} \cdot \boldsymbol{\sigma}, 
\end{eqnarray}
where $\boldsymbol{\sigma}$ is 
a vector consisting of the standard Pauli operators acting on the qubit subspace, by requiring 
the $\pi$ pulse condition $\tau \int_0^1 \Upsilon (s) ds = \pi$. Here, $\tau$ is the run-time 
of the gate and $P_{c}$ is the projection operator onto computational qubit subspace. The 
time-independent complex-valued coupling parameters $\omega_0,\omega_1$ satisfy 
$|\omega_0|^2 + |\omega_1|^2 = 1$ and define the spherical polar angles 
$\theta, \phi$ of the unit vector ${\bf n}$ via the relation $\omega_0/\omega_1 = -e^{-i\phi} 
\tan (\theta/2)$. The gate is holonomic since the Hamiltonian matrix elements 
$\bra{k} e^{i\tau \int_0^1 H(s) ds} H(t) e^{-i\tau \int_0^1 H(s) ds} \ket{l}$, $k,l=0,1$, all 
vanish at any time $t \in [0,\tau]$. Thus, the time evolution is purely geometric and depends 
only on the cyclic evolution path $C_{{\bf n}}$ in the space of two-dimensional subspaces of 
the full three-dimensional Hilbert space of the system, i.e., the Grassmannian manifold $G(3;2)$. 

Now, the fact that the evolution of the computational subspace is purely geometric assures that 
the phase acquired by any input state in the computational subspace is purely geometric.  
Thus, the initial state $\ket{\psi} = a\ket{0} + b\ket{1}$ picks up a purely geometric phase 
$\gamma (\mathcal{C})$ given by the Pancharatnam connection  
\begin{eqnarray}
\gamma (\mathcal{C}) = \arg \bra{\psi}  {\bf n} \cdot \boldsymbol{\sigma} \ket{\psi}
\end{eqnarray}
provided $|\bra{\psi}  {\bf n} \cdot \boldsymbol{\sigma} \ket{\psi}| \neq 0$. By writing 
$\rho = \ket{\psi} \bra{\psi} = \frac{1}{2} (\hat{1} + {\bf r} \cdot \boldsymbol{\sigma})$, with 
$\hat{1}$ and $\boldsymbol{\sigma}$ being the standard Hermitian operator basis on the qubit 
subspace, we obtain 
\begin{eqnarray}
\gamma (\mathcal{C}) = \arg \Tr \left[ \rho U (C_{{\bf n}}) \right] = \arg {\bf n} \cdot {\bf r}  
\end{eqnarray}
which is $0$ or $\pi$ depending on the sign of the scalar product $ {\bf n} \cdot {\bf r}$. 
Note that $\mathcal{C}$ is the path (not necessarily closed) in projective Hilbert space 
$\mathcal{P}$ being isomorphic to the Grassmannian manifold $G(3;1)$. 

All $\ket{\psi}$ with ${\bf r}$ perpendicular to ${\bf n}$ have undefined geometric 
phase $\gamma$. These Bloch vectors form a great circle on the Bloch sphere. States 
below (above) this great circle will pick up $\pi$ ($0$) geometric phase. This is quite 
different from what happens in qubit (spin-$\frac{1}{2}$) precession around a fixed direction 
for which the geometric phase can take any value between $0$ and $2\pi$ depending on 
the angle between the initial Bloch vector and the direction of the precession axis. Furthermore, 
the two cyclic pure states $\rho_{\pm} = \frac{1}{2} (\hat{1} \pm {\bf n} \cdot \boldsymbol{\sigma})$ 
define the computational basis with respect to which the action of $U(C_{{\bf n}})$ defines a 
$\pi$ phase shift ZW gate. 

General ZW gates can be obtained by applying sequentially two pairs of laser pulses. Suppose 
the first pulse pair corresponds to ${\bf n}$ and the second to ${\bf m}$. We obtain
\begin{eqnarray}
U (C_{{\bf m}{\bf n}}) & = & U (C_{{\bf m}}) U (C_{{\bf n}})
\nonumber \\
 & = & {\bf n} \cdot {\bf m} - i \boldsymbol{\sigma} \cdot ({\bf n} \times {\bf m}) 
\label{eq:2pulse}
\end{eqnarray}
for the composite pulse $C_{{\bf m}{\bf n}} = C_{{\bf m}} \ast C_{{\bf n}}$. This SU(2) 
transformation corresponds to a rotation of the qubit with an angle $\vartheta = 
2\arccos \left( {\bf n} \cdot {\bf m} \right)$ around the normal of the plane in 
$\mathbb{R}^3$ spanned by ${\bf n}$ and ${\bf m}$. 

By applying this composite gate to our initial state $\ket{\psi}$ results in the geometric phase 
\begin{eqnarray}
\gamma_g & = & \arg \Tr \left[ \rho U(C_{{\bf m}{\bf n}}) \right] =  
\arg \left( {\bf n} \cdot {\bf m} - i {\bf r} \cdot ({\bf n} \times {\bf m}) \right) 
\nonumber \\ 
 & = & - \tan^{-1} \left( \frac{{\bf r} \cdot ({\bf n} \times {\bf m})}{{\bf n} \cdot {\bf m}} \right) . 
\end{eqnarray}
Provided $|{\bf n} \times {\bf m}| \neq 0$, we notice that the states 
\begin{eqnarray}
\rho_{\pm} = \frac{1}{2} \left( \hat{1} \pm \frac{{\bf n} \times {\bf m}}{ 
|{\bf n} \times {\bf m}|} \cdot \boldsymbol{\sigma} \right) \equiv \ket{\phi_{\pm}} \bra{\phi_{\pm}} 
\end{eqnarray} 
undergo cyclic evolution and pick up the geometric phases 
\begin{eqnarray} 
\gamma_g & = & \mp \tan^{-1} \left( \frac{{\bf n} \times {\bf m}}{|{\bf n} \times {\bf m}|} \cdot 
\frac{{\bf n} \times {\bf m}}{{\bf n} \cdot {\bf m}} \right) = \mp \frac{1}{2} \vartheta . 
\end{eqnarray}
Thus, we obtain the ZW gate 
\begin{eqnarray}
U_g^{\textrm{ZW}} & = & e^{-i\vartheta /2} \ket{\phi_{+}} \bra{\phi_{+}} + 
e^{i\vartheta /2} \ket{\phi_{-}} \bra{\phi_{-}} 
\nonumber \\ 
 & = & \cos \frac{\vartheta}{2} -i \sin \frac{\vartheta}{2} \frac{{\bf n} \times {\bf m}}{ 
|{\bf n} \times {\bf m}|} \cdot \boldsymbol{\sigma} ,
\end{eqnarray}
which is identical to $U(C_{{\bf m}{\bf n}})$ in Eq.~(\ref{eq:2pulse}) since $\cos \frac{\vartheta}{2} = 
{\bf n} \cdot {\bf m}$ and $\sin \frac{\vartheta}{2} = 
|{\bf n} \times {\bf m}|$. 

The main practical advantage with the present implementation of the ZW scheme is that 
by utilizing the third level ($\ket{e}$) the dynamical phases vanish and there is no need 
to invoke compensating operations to make the evolution purely geometric. We note that 
gates with no dynamical phase has been proposed for adiabatic evolution \cite{unanyan04}; 
our proposal can be viewed as a nonadiabatic version of this earlier work.  

To sum up, we have demonstrated that the nonadiabatic non-Abelian geometric gate 
in a three-level $\Lambda$ system proposed in Ref. \cite{sjoqvist12} can alternatively  
be interpreted as a ZW gate based on the same physical scheme. An interesting feature 
of this ZW gate is that no dynamical phases appear and need to be compensated for. 
Moreover, the parameter $\chi$ in ZW two-level scheme \cite{zhu02,zhu03a} is not an 
externally controllable parameter, while in the $\Lambda$ system this parameter is 
controlled by the coupling parameters $\omega_0$ and $\omega_1$.

\section{Adiabatic versus nonadiabatic non-Abelian GQC}
\label{sec:a_vs_na}
A strategy to implement geometric gates is to use adiabatic evolution of energetically 
degenerate subspaces, such as those spanned by the two parameter-dependent dark states 
of a tripod configuration \cite{duan01}. In such a system, the resulting dynamical phases 
are the same for all the states belonging to the subspace and therefore factor out making 
the resulting gate operation purely geometric. The adiabatic approach enables control of 
the evolution by turning the slow parameters around a loop in parameter space so that 
the initial and final Hamiltonians coincide. The geometric gate depends purely on this loop. 

In the nonadiabatic method to realize non-Abelian geometric gates, energy degeneracies 
play no role. Instead, the computational system resides in a subspace of the Hilbert space 
on which the Hamiltonian remains trivial during the execution of the gate. Here, the 
Hamiltonian neither has to return to its initial form nor does it need to evolve slowly as 
long as the initial and final subspaces coincide. The resulting unitary gate is determined 
by the loop performed by the subspace. In its most basic form, nonadiabatic non-Abelian 
geometric quantum computation utilizes the coupling structure of a three-level $\Lambda$ 
system in order to realize a two-dimensional subspace that undergoes a purely geometric 
cyclic evolution \cite{sjoqvist12}. In addition, nonadiabatic schemes based on transitionless 
driving techniques \cite{berry09} have recently been proposed \cite{zhang15,song16}. 

The purpose of this section is to delineate conceptual differences and similarities of adiabatic 
and nonadiabatic non-Abelian geometric quantum computation. 

Let the computational space $\mathcal{M}_c$ be a proper subspace of Hilbert space.  
Let $P_c$ and $\tilde{H}(t)$ be the corresponding projection operator and Hamiltonian, 
respectively. Assume that the run-time of the gate is $\tau$. We may rescale 
the time parameter $t \rightarrow s=t/\tau$ so that $s\in [0,1]$ for the full gate 
performance, which implies that the time evolution 
operator can be written as 
\begin{eqnarray}
U_{\tau} (s,0) = {\bf T} e^{-i \tau \int_0^s H(s') ds'} ,
\end{eqnarray}
${\bf T}$ being time-ordering and $\tilde{H}(t) \rightarrow H(s) = \tilde{H} (s\tau)$. 

First, we describe how geometric gates can be realized by using adiabatic evolution. 
The adiabatic theorem states that \cite{messiah62}
\begin{eqnarray}
\lim_{\tau \rightarrow \infty} U_{\tau} (s,0) P_n (0) = 
P_n (s) \lim_{\tau \rightarrow \infty} U_{\tau} (s,0) , 
\end{eqnarray}
where $P_n (s)$ is an eigenprojector of $H(s)$ associated with the energy eigenvalue 
$\epsilon_n (s)$. Here, we have assumed that $\dot{P}_n (s)$ and $\ddot{P}_n (s)$ 
are well-defined and piecewise continuous, and that $\epsilon_n (s)$ remains distinct 
over $s\in [0,1]$. In practice, adiabatic evolution is enforced by slowly varying some 
experimental control parameters $\boldsymbol{\omega}$ (such as the phases and 
amplitudes of a set of laser beams) around a loop $C_p: [0,1] \ni s \mapsto 
\boldsymbol{\omega} (s)$, $\boldsymbol{\omega} (1) = \boldsymbol{\omega} (0)$, 
in parameter space. 

Now, assume that $P_n (s)$ has constant rank $\geq 2$ and set $P_c = P_n (0)$. 
Thus, we identify the computational space $\mathcal{M}_c$ with the initial eigenprojector 
associated with the energy $\epsilon_n (0)$. Adiabatic geometric gates $U(C_p)$ acting 
on $\mathcal{M}_c$ are realized in the $\tau \rightarrow \infty$ limit for loops $C_p$ 
in parameter space. One finds  
\begin{eqnarray}
 U(C_p) & = & \lim_{\tau \rightarrow \infty} e^{i\tau \int_0^{1} \epsilon_n (s) ds} 
P_c U_{\tau} (1,0) P_c 
\nonumber \\ 
 & = & \sum_{kl} \left( {\bf P}_p e^{i\oint_{C_p} {\bf A} (\boldsymbol{\omega}) 
\cdot d\boldsymbol{\omega}} \right)_{kl} 
\nonumber \\ 
 & & \times \ket{\varphi_k (\boldsymbol{\omega} (0))} \bra{\varphi_l (\boldsymbol{\omega} (0))}
\end{eqnarray}
where ${\bf P}_p$ is path ordering in parameter space and ${\bf A}_{kl} = 
i\bra{\varphi_k (\boldsymbol{\omega})} \nabla_{\boldsymbol{\omega}} \ket{\varphi_l 
(\boldsymbol{\omega})}$ is the matrix-valued Wilczek-Zee vector potential \cite{wilczek84}. 
Thus, for large but finite $\tau$, we have 
\begin{eqnarray} 
P_c U_{\tau} (1,0) P_c \approx e^{-i\tau \int_0^{1} \epsilon_n (s) ds} U(C_p) . 
\label{eq:ahqc}
\end{eqnarray}  
This demonstrates that in the adiabatic regime the dynamical phase factor 
$e^{-i\tau \int_0^{1} \epsilon_n (s) ds}$ approximately factors out and the nontrivial 
action of the evolution on the computational subspace coincides with the non-Abelian 
geometric phase $U(C_p)$. 

Let us next turn to the nonadiabatic case. For Schr\"odinger evolution 
\begin{eqnarray} 
P(0) = P_c \mapsto P(s) & = & U_{\tau}(s,0) P_c U_{\tau}^{\dagger} (s,0) 
\nonumber  \\ 
 & = & \sum_{k} \ket{\psi_k (s)} \bra{\psi_k (s)}, 
\end{eqnarray} 
one obtains \cite{anandan88} 
\begin{eqnarray} 
P(s) U_{\tau}(s,0) P(s) & = & \sum_{kl} \left( {\bf P}_{g} e^{-i\tau \int_0^s {\bf \mathcal{K}}(s') ds' + 
i \int_0^s {\bf \mathcal{A}} (s') ds'} \right)_{kl} 
\nonumber \\ 
 & & \times \ket{\psi_k (s)} \bra{\psi_l (s)} , 
\end{eqnarray}
where \begin{eqnarray}
\left( {\bf \mathcal{K}} \right)_{kl} & = & \bra{\psi_k (s)} H(s) \ket{\psi_l (s)}, 
\nonumber \\ 
\left( {\bf \mathcal{A}} \right)_{kl} & = & i\langle \psi_k (s) \ket{\dot{\psi}_l (s)}, 
\end{eqnarray} 
and ${\bf P}_g$ is path ordering in the Grassmannian $G(N;K)$, i.e., the space of 
$K$-dimensional subspaces (assuming $P(s)$ has fixed rank $K$) of an $N$-dimensional 
Hilbert space. A general condition for making the gate purely geometric is 
\begin{eqnarray}
\left( {\bf \mathcal{K}} \right)_{kl} = \eta (s) \delta_{kl}  
\label{eq:nonadpt}
\label{eq:nageom}
\end{eqnarray}
with $\eta (s)$ an arbitrary real-valued function. Under this condition, one obtains 
\begin{eqnarray}
P_c U_{\tau} (1,0) P_c & = & e^{-i \tau \int_0^1 \eta (s) ds} 
\sum_{kl} \left( {\bf P}_g e^{i\oint_{C_g} {\bf \mathcal{A}}} \right)_{kl} 
\nonumber \\ 
 & & \times \ket{\psi_k (0)} \bra{\psi_l (0)}  
\label{eq:nahqc}
\end{eqnarray}
provided there exists a $\tau$ such that $P(1)=P(0)=P_c$ (cyclic evolution).  One should 
note that $\mathcal{A}$ is defined for a smooth single-valued family of bases 
$\{ \ket{\psi_k (s)} \}_{k=1}^K$, i.e., $\ket{\psi_k (s)}$ is differentiable and 
$\ket{\psi_k (1)} = \ket{\psi_k (0)}$, $\forall k$. The nontrivial part  
\begin{eqnarray}
U(C_g)= {\bf P}_g e^{i\oint_{C_g} {\bf \mathcal{A}}}
\end{eqnarray}
describes a purely geometric action on $\mathcal{M}_c$. Explicitly, $U(C_g)$ is the holonomy 
of the loop $C_{g}$ based at the computational space $\mathcal{M}_c$ in the Grassmannian 
$G(N;K)$.

Let us now identify conceptual differences and similarities between the adiabatic 
and non-adiabatic approaches to implement non-Abelian geometric gates.  
\begin{itemize}
\item {\it Role of the run-time $\tau$.} In the adiabatic case, $\tau$'s role is to factor 
out the dynamical phase and make the nontrivial action of the gate purely geometric 
\cite{simon83}. This is achieved in the adiabatic limit where $\tau \rightarrow \infty$. 
In the nonadiabatic case, on the other hand, $\tau$'s role is to make sure the evolution 
is cyclic. In other words, $\tau$ is finite and can even be short compared to the 
intrinsic time scale related to the energy shifts of the Hamiltonian, in the 
nonadiabatic scenario.  
\item {\it Exactness of the gates.} The adiabatic gate in Eq.~(\ref{eq:ahqc}) becomes exact 
only in the mathematical limit where $\tau$ tends to infinity. Since all experiments involve 
finite $\tau$, it therefore follows that adiabatic geometric gate can never be exact; there 
will be nonadiabatic dynamical corrections to the gate that can in principle be made arbitrarily 
small but never precisely zero \cite{florio06}. On the other hand, for a given Hamiltonian 
satisfying the geometry condition Eq.~(\ref{eq:nageom}), the resulting 
nonadiabatic geometric gate is exact and can be implemented for a finite $\tau$. 
\item {\it Role of the parameters $\boldsymbol{\omega}$.} While the geometric phase 
is induced by slow changes of physical control parameters $\boldsymbol{\omega}$ in the 
adiabatic case, these parameters play a passive role in the nonadiabatic version. They may 
even be kept fixed during the execution of a nonadiabatic gate. 
\item {\it Interpretation of the loops $C_p$ and $C_g$.} The adiabatic loop $C_p$ is traced 
out in a space of slow parameters. The nonadiabatic loop $C_g$ is traced out in a Grassmannian. 
Thus, these two loops are traced out in different types of spaces. However, one may equally 
well associate the adiabatic geometric phase with the loop traversed by the energy eigensubspace 
in the corresponding Grassmannian \cite{fujii00}. Thus, on a fundamental level, all geometric 
gates, no matter if they arise in adiabatic or nonadiabatic evolution, depend on paths in a 
Grassmannian. 
\end{itemize}

In order to clarify further the above points, let us now examine the main approaches to 
adiabatic and nonadiabatic geometric quantum computation, viz., the tripod and $\Lambda$ 
setting, respectively, see Fig.~\ref{fig:tripodlambda}. The tripod configuration consists of 
three `ground state' energy levels 
$\ket{0}, \ket{1},\ket{a}$ coupled by three laser fields to one and the same excited state 
$\ket{e}$; the $\Lambda$ configuration is simply the tripod with $\ket{a}$ removed. 
The detailed nature of the underlying physical system is not important as long as it obeys this 
structure. It can, e.g., be a trapped ion addressed by lasers fields \cite{toyoda13}, it can 
be a transmon qubit \cite{abdumalikov13} or nitrogen-Vacancy center in diamond 
\cite{arroyo-camejo14,zu14} driven by microwave fields. By employing the rotating wave 
approximation (RWA) in the interaction picture, we obtain the tripod Hamiltonian
\begin{eqnarray}
H^{{\textrm{tripod}}} & = & \Delta_0 \ket{0} \bra{0} + \Delta_1 \ket{1} \bra{1} + 
\Delta_a \ket{a} \bra{a} 
\nonumber \\ 
 & & + \Upsilon \Big( \omega_0\ket{e} \bra{0} + \omega_1 \ket{e} \bra{1} 
\nonumber \\ 
 & & + \omega_a \ket{e} \bra{a} + {\textrm{h.c.}} \Big)
\end{eqnarray}
and the $\Lambda$ Hamiltonian 
\begin{eqnarray}
H^{\Lambda} & = & \Delta_0 \ket{0} \bra{0} + \Delta_1 \ket{1} \bra{1} 
\nonumber \\ 
 & & + \Upsilon \Big( \omega_0 \ket{e} \bra{0} + \omega_1 \ket{e} \bra{1} + 
{\textrm{h.c.}} \Big) 
\end{eqnarray}
with time-independent detunings $\Delta_j = 2\pi \nu_j - \omega_{je}$, $\nu_j$ and 
$\omega_{je}$ being the field frequencies and energy spacings, respectively, and $\omega_j$ 
being complex-valued parameters describing the phase and amplitude of the fields. In both 
cases, the computational subspace is $\mathcal{M}_c={\textrm{Span}} \{ \ket{0} , \ket{1} \}$. 
We further assume that $\sum_j |\omega_j|^2 = 1$, which means that $\Upsilon$ measures 
the overall strength of the laser-atom interaction.   

Let us first see how the adiabatic tripod scheme works. Here, we first look for restrictions 
on the parameters that generate a degenerate pair of energy eigenstates of the form 
$c_0\ket{0} + c_1\ket{1} + c_a\ket{a}$. These are called dark states as they do not involve 
the potentially unstable excited state $\ket{e}$. Given this form, the eigenvalue equation 
for $H^{{\textrm{tripod}}}$ gives   
\begin{eqnarray}
\omega_0 c_0 +\omega_1 c_1 +\omega_a c_a & = & 0 ,  
\nonumber \\ 
\Delta_0 c_0 & = & \epsilon c_0 , 
\nonumber \\ 
\Delta_1 c_1 & = & \epsilon c_1 , 
\nonumber \\ 
\Delta_a c_a & = & \epsilon c_a , 
\end{eqnarray} 
$\epsilon$ being the energy eigenvalue of the dark state subspace. These equations 
have precisely two solution if and only if 
\begin{eqnarray} 
\Delta_0 = \Delta_1 = \Delta_a \equiv \epsilon. 
\label{eq:adetuning}
\end{eqnarray}
Thus, there is a degenerate pair of dark energy eigenstates $\ket{D_0 (\boldsymbol{\omega})}$ 
and $\ket{D_1 (\boldsymbol{\omega})}$, for all $\boldsymbol{\omega} = 
(\omega_0,\omega_1,\omega_a)$, with energy $\epsilon$ being the common detuning of 
the three laser fields. In addition, there are two nondegenerate bright states $\ket{B_{\pm} 
(\boldsymbol{\omega})}$ with energies $\frac{1}{2} \left( \epsilon \pm \sqrt{\epsilon^2 + 
4\Upsilon^2} \right)$. With $P_d (\boldsymbol{\omega}) = \ket{D_0 (\boldsymbol{\omega})} 
\bra{D_0 (\boldsymbol{\omega})} + \ket{D_1 (\boldsymbol{\omega})} 
\bra{D_1 (\boldsymbol{\omega})}$, we can thus write 
\begin{eqnarray}
 & & H^{{\textrm{tripod}}} = \epsilon P_d (\boldsymbol{\omega}) 
\nonumber \\ 
 & & + \frac{1}{2} \left( \epsilon + \sqrt{\epsilon^2 + 4\Upsilon^2} \right) 
\ket{B_+ (\boldsymbol{\omega})} \bra{B_+ (\boldsymbol{\omega})} 
\nonumber \\ 
 & & + \frac{1}{2} \left( \epsilon - \sqrt{\epsilon^2 + 4\Upsilon^2} 
\right) \ket{B_- (\boldsymbol{\omega})} \bra{B_- (\boldsymbol{\omega})}
\end{eqnarray}
when Eq. (\ref{eq:adetuning}) holds. 

Now we assume that $\boldsymbol{\omega} = \boldsymbol{\omega} (s)$ varies around 
a loop $C_p$ in parameter space and that $\Upsilon = \Upsilon (s)$ is nonzero on $s\in [0,1]$. 
In the adiabatic regime, $\tau$ is so large that transitions between the dark subspace and the 
two bright states become negligible. The condition for this is 
\begin{eqnarray} 
\tau \gg \frac{1}{\min_{s\in [0,1]} \left\{ \frac{1}{2} 
\left( \epsilon - \sqrt{\epsilon^2 + 4\Upsilon (s)^2} \right) \right\} } , 
\end{eqnarray}
i.e., $\tau$ should be large compared to the inverse of the minimal energy gap. 
When this condition is satisfied, the loop $C_p$ in the space of slowly changing 
$\boldsymbol{\omega}$ approximately determines the nontrivial action of the 
time evolution operator and would be a gate acting on $\mathcal{M}_c$ provided 
$P_d (\boldsymbol{\omega} (0)) = P_{c} = \ket{0}\bra{0} + \ket{1}\bra{1}$, which is achieved 
by choosing $\boldsymbol{\omega} (0) = (0,0,1)$. We note that the space of all dark 
subspaces is the Grassmanian manifold $G(3;2)$, i.e., the space of the two-dimensional 
subspaces ${\textrm{Span}} \{ \ket{D_0 (\boldsymbol{\omega})} , \ket{D_1 (\boldsymbol{\omega})} \}$ 
of the three-dimensional vector space ${\textrm{Span}} \{ \ket{0},\ket{1},\ket{a}\}$ 
(for a proof, see Appendix). Hence, the loop $C_p$ in the space of slow parameters 
induces a loop in $G(3;2)$ initiated at the computational space $\mathcal{M}_c={\textrm{Span}} 
\{ \ket{0},\ket{1}\}$. The gate $U(C_p)$ is the non-Abelian geometric phase associated with 
this loop in $G(3;2)$.

Next, we turn to the nonadiabatic case. We look for restrictions on the parameters 
in $H^{\Lambda}$ such that Eq. (\ref{eq:nonadpt}) is satisfied. This is equivalent to 
solving the nonlinear equations  
\begin{eqnarray}
 & \bra{k} \left( {\bf T} e^{-i \tau \int_0^s H^{\Lambda} (s') ds'} \right)^{\dagger} H^{\Lambda} (s) 
{\bf T} e^{-i \tau \int_0^s H^{\Lambda} (s') ds'} \ket{l} 
\nonumber \\ 
 & = \eta (s) \delta_{kl} , \ k,l=0,1, \ \forall s\in [0,1] . 
\end{eqnarray}
A nontrivial solution can be found by assuming that $\Upsilon,\omega_0,\omega_1$ are 
$s$-independent during $s\in [0,1]$, resulting in the simplified equations 
\begin{eqnarray} 
\bra{k} H^{\Lambda} \ket{l} & = & \Delta_l \delta_{kl} = \eta \delta_{kl} , 
\end{eqnarray} 
which implies 
\begin{eqnarray}
\Delta_0 = \Delta_1 = \eta.
\end{eqnarray}
It can be shown that this choice is sufficient for the realization of a universal purely geometric 
single-qubit gate acting on the computational subspace $\mathcal{M}_c=\textrm{Span} 
\{ \ket{0},\ket{1} \}$ provided the run-time $\tau$ satisfies \cite{xu15,sjoqvist16} 
\begin{eqnarray} 
\tau = \frac{2\pi}{\sqrt{\eta^2+4\Upsilon^2}} . 
\end{eqnarray}
The resulting geometric gate is exact under the assumption that the RWA is valid 
\cite{spiegelberg13}. 

To sum up, geometric quantum computation based on non-Abelian geometric phases 
can be implemented by using adiabatic or nonadiabatic evolution. While the former 
relies on an infinite run-time, and can therefore never be exact without loosing its 
geometric character, the latter is exact and can be implemented at high speed. 
Conceptually, both types of geometric phases can be associated with loops in a 
Grassmann manifold with start- and end-point coinciding with the computational 
subspace. While the Hamiltonian parameters are used to steer a energetically 
degenerate subspace around a loop in the adiabatic approach, these parameters 
play a passive role in the nonadiabatic case and can even be kept fixed during the 
execution of the gate.   
 
\section{Conclusions}
\label{ref:conclusion}
Geometric quantum computation is an approach to implement quantum gates by 
using different types of geometric phases. These phases can be Abelian or non-Abelian, 
which can be realized in adiabatic or nonadiabatic evolution. The purpose of the 
present work has been to shed light on some conceptual issues related to these 
different forms of geometric gates.  

In Sec.~\ref{sec:dp}, we have examined under what circumstances a gate can be said to be 
geometric. A condition for this is that the considered gate contains no dynamical phase effects. 
We have argued that there exist two different types of dynamical phases, where one is 
of global nature and therefore has no observable consequences, while the other one is of 
relative nature and directly influences the effect of the gate so that it must be removed. 
In other words, one may allow for dynamical phases in geometric gates, provided these 
phases are of global rather than relative nature.  

It is known that all-geometric quantum computation can be implemented by 
using both Abelian and non-Abelian geometric phases. Since these conceptually 
very different types of phases apparently achieve exactly the same, one may ask whether 
there is any relation between them. In Sec.~\ref{sec:zwhqc}, we have shown that this can 
indeed be the case, by demonstrating that the Abelian and non-Abelian approaches 
give rise to the same set of gates in a three-level $\Lambda$ system. 

Geometric gates can be characterized by whether the underlying evolution is adiabatic 
or nonadiabatic. In Sec. \ref{sec:a_vs_na}, we have discussed differences and similarities  
between these two types of gates, in the case of non-Abelian geometric phases. Adiabatic 
and nonadiabatic non-Abelian geometric gates are similar in that they both are based on 
matrix-valued geometric phases and that they can interpreted in terms of loops in a 
Grassmann manifold. The main differences concern the role of the run-time 
and control parameters. In the adiabatic case, the run-time is used to factor out the 
dynamical phase and the control parameters play an active role to move the degenerate 
energy subspace. In the nonadiabatic case, on the other hand, the run-time is used to 
ensure the evolution is cyclic. The control parameters play a passive role and can even 
be kept fixed during the execution of such nonadiabatic gates.  

The existence of a wide range of conceptually very different schemes implies that geometric 
quantum computation can be implemented in many different physical systems. It provides 
a rich tool-box for addressing different types of errors that occur in different quantum gate 
architectures. For instance, the adiabatic schemes can be used in cases where parameter 
noise is present, while nonadiabatic schemes can be used in cases where decoherence is 
present by reducing the exposure time. Thus, geometric quantum computation offers  
a conceptual framework that can be used as a guiding tool in the realization of quantum 
computers. 
\section*{Appendix}
\label{sec:appendix}
We prove that the space of all dark subspaces of the tripod system is $G(3;2)$. We do 
this by demonstrating that for any $\ket{\psi} \in \textrm{Span} \{ \ket{0}, \ket{1}, \ket{a} \}$ there exists $\boldsymbol{\omega}$ such that 
\begin{eqnarray}
P_d (\boldsymbol{\omega}) \ket{\psi} = 0.
\label{eq:condition}
\end{eqnarray} 
By using the linear 
independence of the two dark states $\ket{D_0 (\boldsymbol{\omega})},
\ket{D_1 (\boldsymbol{\omega})}$, it follows that Eq.~(\ref{eq:condition}) is equivalent to 
\begin{eqnarray}
\langle D_j (\boldsymbol{\omega}) \ket{\psi} = 0, \ \ \  j=0,1, 
\label{eq:altcondition}
\end{eqnarray}
for $\ket{\psi} = \lambda_0 \ket{0} + \lambda_1 \ket{1} + \lambda_a \ket{a}$ with arbitrary 
complex-valued $\lambda_0,\lambda_1,\lambda_a$ such that $|\lambda_0|^2 + 
|\lambda_1|^2 + |\lambda_a|^2 \neq 0$. By using the explicit form of the two dark 
states (parameterization taken from Ref.~\cite{ruseckas05}), we find 
\begin{eqnarray}
 & \sin \phi  e^{-iS_{31}} \lambda_0 - \cos \phi  e^{-iS_{32}} \lambda_1 = 0 , 
\nonumber \\ 
 & \cos \theta \cos \phi  e^{-iS_{31}} \lambda_0 + \cos \theta \sin \phi  e^{-iS_{32}} \lambda_1 
\nonumber \\ 
 & - \sin \theta \lambda_a = 0 , 
\label{eq:arbitrary}
\end{eqnarray}
where $\omega_0 = \sin \theta \cos \phi e^{iS_1}$, $\omega_1 = \sin \theta \sin \phi e^{iS_2}$, 
$\omega_a = \cos \theta e^{iS_3}$, and $S_{kl} = S_k - S_l$. 

Assume first that $\lambda_0 \neq 0$ and define $z_1=\lambda_1/\lambda_0, 
z_a=\lambda_a/\lambda_0$. We find 
\begin{eqnarray} 
\tan \phi e^{-iS_{21}} & = & z_1, 
\nonumber \\ 
\cot \theta e^{-iS_{31}} & = & z_a \cos \phi .
\label{eq:c0}
\end{eqnarray}
This can be solved for all $z_1,z_a$ since $\theta,\phi,S_{21},S_{31}$ are independent 
variables. Explicitly, one finds $\phi = \tan^{-1} |z_1|$, $S_{21} = -\arg z_1$, 
$\theta = \cot^{-1} \left[ |z_a|/ \sqrt{1+|z_1|^2}\right]$, and $S_{31} = -\arg z_a$. 
Next, we assume that $\lambda_0 = 0$ but $\lambda_1 \neq 0$, and define 
$\tilde{z}_a = \lambda_a/\lambda_1$. We find 
\begin{eqnarray} 
\phi & = & \frac{\pi}{2} , 
\nonumber \\ 
\cot \theta e^{-iS_{32}} & = & \tilde{z}_a 
\label{eq:c1}
\end{eqnarray}
with solution $\theta = \cot^{-1} | \tilde{z}_a|$ and $S_{32} = -\arg \tilde{z}_a$. Finally, if 
$\lambda_0 = \lambda_1=0$, then $\theta = 0$ solves  Eq.~(\ref{eq:arbitrary}). 
\begin{acknowledgements}
E.S. acknowledges financial support from the Swedish Research Council (VR) through 
Grant No. D0413201. V.A.M. acknowledges support from the Department of Mathematics 
at University of Isfahan (Iran). C.M.C. is supported by Department of Physics and Electrical 
Engineering at Linnaeus University (Sweden) and by the Swedish Research Council (VR) 
through Grant No. 621-2014-4785. 
\end{acknowledgements}

\end{document}